\documentclass[utf8]{preprint}

\usepackage{url,hyperref,lineno,microtype}
\usepackage[onehalfspacing]{setspace}

\newcommand{\ie}{\textit{i.e.\@}} 
\newcommand{\eg}{\textit{e.g.\@}}

\def\keyFont{\fontsize{8}{11}\helveticabold }

\def\Authors{Remy Gardier\,$^{1,*}$, Juan Luis Villarreal Haro\,$^{1}$, Erick J Canales-Rodríguez\,$^{1}$, Ileana O. Jelescu\,$^{2,3}$, Gabriel Girard\,$^{1,2, 4}$,  Jonathan Rafael-Patino\,$^{2,1}$, Jean-Philippe Thiran\,$^{1, 2, 4}$}

\def\Address{$^{1}$Signal Processing Laboratory (LTS5), École Polytechnique Fédérale de Lausanne (EPFL), Lausanne, Switzerland\\
$^{2}$Radiology Department, Centre Hospitalier Universitaire Vaudois (CHUV) and University of Lausanne (UNIL), Lausanne, Switzerland\\
$^{3}$School of Biology and Medicine, University of Lausanne (UNIL), Lausanne, Switzerland\\
$^{4}$CIBM Center for Biomedical Imaging, Switzerland}

\def\corrAuthor{Remy Gardier, Signal Processing Laboratory (LTS5), Station 11, CH-1015 Lausanne,
Switzerland}

\def\corrEmail{remy.gardier@epfl.ch}

\def\title{Microstructure estimation from diffusion-MRI: Compartmentalized models in permeable cellular tissue}

\begin{document}
\onecolumn
    \sffamily
   \parindent0pt
   
    {\vspace{5cm}  \LARGE  \bfseries \title}
    
    {\vspace{0.5cm} \bfseries \Authors}
    
    {\vspace{0.5cm}  \scriptsize \itshape \Address{}} 
    
    {\vspace{0.5cm}   \footnotesize Correspondence*: \\ \corrAuthor \\ \corrEmail}

\begin{abstract}

\section{}

Diffusion-weighted magnetic resonance imaging (DW-MRI) is used to characterize brain tissue microstructure employing tissue-specific biophysical models. A current limitation, however, is that most of the proposed models are based on the assumption of negligible water exchange between the intra- and extracellular compartments, which might not be valid in various brain tissues, including unmyelinated axons, gray matter, and tumors. The purpose of this work is to quantify the effect of membrane permeability on the estimates of two popular models neglecting exchange, and compare their performance with a model including exchange. To this aim, DW-MRI experiments were performed in controlled environments with Monte-Carlo simulations. The DW-MRI signals were generated in numerical substrates mimicking biological tissue made of spherical cells with permeable membranes like cancerous tissue or the brain gray matter. From these signals, the substrates properties were estimated using SANDI and VERDICT, the two compartment-based models neglecting exchange, and CEXI, a new model which includes exchange. Our results show that, in cellular permeable tissue, the model with exchange outperformed models without exchange in the estimation of the tissue properties by providing more stable estimates of cell size, intracellular volume fraction and extracellular diffusion coefficient. Moreover, the model with exchange estimated accurately the exchange time in the range of permeability reported for cellular tissue. Finally, the simulations performed in this work showed that the exchange between the intracellular and the extracellular space cannot be neglected in permeable tissue with a conventional PGSE sequence, to obtain accurate estimates. Consequently, existing compartmentalized models of impermeable tissue cannot be used for microstructure estimation of cellular permeable tissue.

\tiny
 \keyFont{ \section{Keywords:} Diffusion MRI, Microstructure, Permeability, Monte-Carlo Simulations, Compartmentalized model, Tumor, Exchange, Time dependence} 
\end{abstract}

\section{Introduction}

Diffusion-weighted magnetic resonance imaging (DW-MRI) is a non-invasive imaging technique sensitive to the displacement of water molecules in brain tissue. Over the last decades, it has been used to characterize brain tissue microstructure. On one side, mathematical expressions of the diffusion signal, such as Diffusion Tensor Imaging ($DTI$) \citep{Basser1994} and Diffusion Kurtosis Imaging ($DKI$) \citep{Jensen2005} were proposed to estimate the main diffusivities, the Apparent Diffusion Coefficient ($ADC$), and the Apparent Diffusion Kurtosis ($ADK$), respectively. These metrics have been helpful to complement the diagnosis of a variety of diseases both inside the brain, such as Parkinson’s \citep{Wang2011} and tumor grading \citep{VanCauter2012, Roethke2015}, and outside the brain \citep{Rosenkrantz2015}.

On the other side, various biophysical models were develop to estimate specific tissue microstructure properties. These models group the biological entities into compartments with similar contribution to the diffusion signal, and differ essentially by the number of compartments and the targeted microstructure features. For example, the mean axon diameter \citep{Alexander2010}, the diameter distribution \citep{Assaf2008, Assaf2005, Assaf2004}, the fiber dispersion, and the axonal density \citep{Zhang2012, Jespersen2007, Tariq2016, Kaden2016, Novikov2018} were the first metrics used to characterize the brain white matter microstructure. More recently, the Standard Model \citep{Novikov2019}, a two-compartment model,  was proposed to unite these models in the long-time diffusion limit (assuming Gaussian diffusion), where axons are modeled as impermeable cylinders with a zero radius (\ie, sticks), and the extra-axonal space as an anisotropic Gaussian compartment.

One of the challenges of biophysical models is their specificity to one particular tissue, and each tissue type requires rethinking the optimal model and assumptions that best capture its features. For example, in gray matter with signals heavily diffusion-weighted (at high $b$ value), the stick assumption does not fit the signal accurately \citep{McKinnon2017, Veraart2019, Veraart2020}, and some studies suggested that the intra-soma diffusion requires a specific compartment \citep{Palombo2018a, Palombo2018b}. Additionally, the assumption of negligible water exchange between compartments in most models might not be valid in unmyelinated microstructure, urging for specific models of microstructure in permeable tissue \citep{Aggarwal2020, Reynaud2017, Olesen2022, Jelescu2020b}. Two relevant examples of tissue with unmyelinated microstructure are the brain gray matter and tumors. Again, both required their own biological model because the neurites in gray matter cannot be neglected, and the spherical structure (the somas in gray matter and the cells in cancerous tissue) differ in size (in ranges 5-20 $\mu m$ and 1-5 $\mu m$, respectively), density (10-20\%, at least 40\%), and exchange time \citep{Palombo2021, Yang2018, Veraart2020, Olesen2022, Jelescu2020a, Zhang2021, Aggarwal2020, Zhao2008, Reynaud2016}. As of now, permeable tissue have been characterized with biophysical models that either neglect the effect of water exchange on the signal by employing acquisition sequences with short diffusion times \citep{Reynaud2016, Palombo2020a, Panagiotaki2014}, or model the exchange based on the two-compartment model of exchange proposed by \cite{Karger1985} (in gray matter \citep{jelescu2022, Olesen2022} and cellular tissues \citep{Karunanithy2019}).

To identify the best microstructure model for a given tissue, and to design the optimal DW-MRI acquisition protocol, it is fundamental to evaluate and validate the available models in controlled environments. In this context, Monte-Carlo Diffusion Simulations (MCDS) \citep{Patino2020, Lee2021, Hall2009, Hall2017, Brusini2019} can be used to simulate the diffusion MRI signals in complex substrates. MCDS provides two main benefits 1) the diffusion signal can be generated without assuming an analytical equation, thus eliminating the comparison bias across different methods (introduced when a functional form is assumed), and 2) realistic substrates \citep{Fieremans2010, Abdollahzadeh2019, Lee2020, Andersson2020, Patino2020} can be designed allowing to study various microstructure parameters, including the permeability \citep{Lee2021, Hall2017}, in a controlled manner.

In this work, the performances of compartmentalized models excluding and including water exchange between compartments were studied in permeable tissue mimicking tumors cells by means of MCDS in complex substrates. More specifically, the work focused on the influence of non-neglectable membrane permeability on microstructure estimation, \ie~ the cell sizes, the extracellular diffusion coefficient, and the intracellular volume fraction, with two compartmentalized models neglecting exchange (SANDI \citep{Palombo2020a} and VERDICT \citep{Panagiotaki2014}) and one model including exchange (CEXI, a counterpart of \citep{jelescu2022} adapted for spherical cells). Both SANDI \citep{Palombo2020a} and VERDICT (Vascular, Extracellular and Restricted Diffusion for Cytometry in Tumors) \citep{Panagiotaki2014} are three-compartment models developed under the assumption of negligible exchange, designed for gray matter and cancerous tissue respectively. Conversely, CEXI (Cellular Exchange Imaging) is a two-compartment model of cell microstructure that includes the exchange between spherical cells and extracellular space. The rest of this article is organized as follows. These models and the theoretical framework are presented in details in Section ~\ref{sec:theory}, while the conducted experiments are described in Section \ref{sec:experiments}. In the Results and Discussion (i.e., Sections \ref{sec::results} and \ref{sec::discussion}), the relevant findings are briefly described, along with some recommendations for future works. This work's main findings and contributions are addressed in the Conclusion (Section \ref{sec::conclusion}.

By means of Monte-Carlo simulations in realistic numerical substrates of tumors, the effects of permeability on microstructure estimation by compartment-based models were investigated. The simulation framework and the choice of the simulation parameters and numerical substrate details are presented in Section~\ref{sec:method_mcdc}. Next, Section ~\ref{sec:method_comp} outlines the compartmentalised models of microstructure evaluated in this simulation framework. Finally, Section ~\ref{sec:method_validation} and ~\ref{sec:method_model} describe the experiments designed for the simulation framework validation and the compartmentalized models comparison.

\subsection{Theory}\label{sec:theory}
\subsubsection{Realistic Monte-Carlo simulations in a permeable substrate}\label{sec:method_mcdc}

The MC-DC Simulator of~\cite{Patino2020} initially developed for DW-MRI experiments in impermeable substrates was extended for diffusion in tissue with multiple diffusion coefficients $D$ \citep{Szafer1995} and permeability $\kappa$ \citep{Lee2021}. In non-exchanging mediums, diffusion inside different biological structures are assumed to contribute independently to the DW-MRI signal. Consequently, the intracellular and extracellular signals are usually generated individually, and summed to produce the full signal. In permeable tissue, for which the non-exchanging assumption is not valid, this approach might produce unrealistic signals due to some particles undergoing diffusion properties modifications after crossing the membranes. For this reason, the simulator required the implementation of the multiple-diffusivity features to generate intracellular and extracellular signals simultaneously. In simulations with multiple diffusion coefficients, the time step $\delta t$ and the step length $\delta s$, related by Einstein's equation $\delta s = \sqrt{6D\delta t}$, might not remain both constant during the simulation. Therefore, the simulator kept $\delta t$ constant, and updated $\delta s$ for each particle during a step (See supplementary materials for derivations). Once the particles' trajectories were simulated, the signals were generated with a Graphics Processing Unit (GPU) implementation, and a bootstrapping analysis was performed. 

\begin{figure}[h!]
\begin{center}
\includegraphics[width=\textwidth]{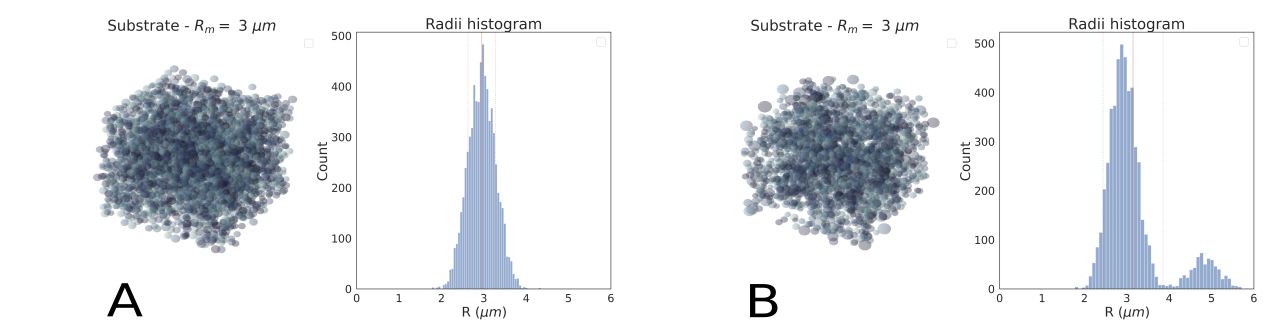}
\end{center}
\caption{\textbf{Numerical substrates.} Example of numerical substrates (Table ~\ref{tab:substrate}) with a mean volume-weighted radius $R \simeq 3\mu m$. (A) The substrate $S_2$ had one population of sphere with radii distributed around the mean $R_s = 3 \mu m$. (B) The substrate $M_2$ was composed of a small and a big sphere populations of mean radius $R_s = 1 \mu m$ and $R_s = 5 \mu m$.  The histograms highlight the difference between the radii distributions. The solid red and dashed lines shows the mean and the standard deviation of $R$, respectively. }\label{fig:substrate_example}
\end{figure}

In addition to the rules of the particles' dynamic, Monte-Carlo simulations require a numerical substrate of the tissue of interest into which the particles diffuse. To mimic tumors, the substrates were settled as isotropic voxels of side-length $100 \mu m$, filled of spheres from one or two populations. Their radii were normally distributed around a mean radius $R_s$. In the single-$R_s$ substrates, the mean radius was equal to $\{2; 3; 4; 5 \}\mu m$, whereas the multiple-$R_s$ was a mixture of two radii equal to $\{1, 3; 1, 5; 3, 5 \} \mu m$. 
The sphere packing were generated with a two-stage process. The sphere were initialized randomly in the voxel without overlapping, from the largest to the smallest radius. From this initial configuration, the packing was optimized \citep{Baranau2017} to the desired intracellular volume fraction $ICVF$ of $0.65$ for all substrates. In the two-population substrates, the volume fractions of the largest and smallest spheres were 40\% and 60\% of the $ICVF$, respectively. Because the spheres of a substrate had different radii, the substrates were additionally characterized by their volume-weighted mean cell radius $R$

\begin{equation}
    R = \sqrt[3]{\frac{\sum_{i=1}^{N_s}{R_{s,i}^3}} {N_s}},
\end{equation}

where $N_s$ is the number of spheres and $R_{s,i}$ is the radius of the $i^{th}$ sphere. Table~\ref{tab:substrate} summarizes the structural properties of the numerical substrates used in the experiments, and Fig.\ref{fig:substrate_example} shows illustrative examples of substrates $S_2$ (Fig. \ref{fig:substrate_example}A) and $M_2$ (Fig. \ref{fig:substrate_example}B) with a volume-weighted mean radius $R$ close to $3\mu m$. The left subplots shows the location of the spheres in the substrate, while the right subplots show the sphere radii distribution for each substrate.

In addition to structural properties, particle trajectories also depend on the biological properties of the substrates. In permeable tissue, these are the intracellular $D_{i, 0}$ and extracellular $D_{e, 0}$ diffusion coefficients and the membrane permeability $\kappa$. Their ranges in the simulations were chosen based on previously reported values for tumors \citep{Mukherjee2016} (Table~\ref{tab:experiment}). In the validation experiment (Section~\ref{sec:method_validation}), the membrane permeability $\kappa$ ranged from 0 to $ 100\frac{\mu m}{s}$, the intracellular diffusion coefficient $D_{i, 0}$ was set to  $1$ or $2 \frac{m^2}{s}$, and the extracellular diffusion coefficient $D_{e, 0}$ was set to $0.5, 1$ or $2\frac{m^2}{s}$. In the model comparison experiment (Section ~\ref{sec:method_model}),  $\kappa$ was set 0, 10, 25 or $ 50 \frac{\mu m}{s}$, $D_{i, 0}$ was fixed to  $2 \frac{m^2}{s}$, and $D_{e, 0}$ was set to $0.5, 1$ or $2\frac{m^2}{s}$.

Finally, the DW-MRI signals were generated with PGSE sequences, for which the parameters are described in Section~\ref{sec:method_comp} and summarized in Table~\ref{tab:pgse}. In protocols with fixed $b$ values and diffusion times, the gradient strength $g$ was calculated by the relation $b = (\gamma \delta g)^2 (\Delta- \frac{\delta}{3})$, where $\gamma$ is the gyromagnetic ratio. Signals were generated in $N$ uniformly distributed directions for each combination of the parameters even if the substrates were isotropic by design. 

\subsubsection{Compartmentalised models of microstructure}\label{sec:method_comp}
\paragraph{SANDI}
The three-compartment model SANDI \citep{Palombo2020a} was developed to model gray matter microstructure properties under the assumption of negligible exchange, which was validated at short diffusion time ($\Delta < 20 ms$  originally). SANDI models the neurites and the somas as sticks and spheres with diffusion $D_{i,n}$ and $D_{i,s}$ respectively, and the extracellular diffusion is assumed Gaussian and isotropic with diffusivity $D_{ex}$. The signal fractions $(f_ {ex}, f_ {n}, f_ {s})$ of each compartment, the mean radius $R_s$ of the soma and the diffusion coefficient of the neurites and the extracellular space were estimated from the normalized powder-averaged signal $\tilde{S}_{SANDI}$

\begin{equation}
    \tilde{S}_{SANDI}(b) = \left(1-f_ {ex}\right) \left(f_n \tilde{S_n} + f_s \tilde{S_s} \right) + f_{ex} \tilde{S_{ex}}.
\end{equation}

Mathematical formulation of the powder-averaged signal of the extracellular $\tilde{S}_{ex}$, the neurites $\tilde{S}_n$ and the soma $\tilde{S}_s$ of radius $R_s$ compartments were derived in~\cite{Palombo2020a} (Eq. 6-8).

In the original paper, the intra-soma diffusion coefficient $D_{i,s}$ was fixed to $3 \frac{\mu m^2}{ms}$. 
The results of the experiments described in Section~\ref{sec:method_model} were estimated with the open-source Accelerated Microstructure Imaging via Convex Optimization (AMICO) implementation of SANDI \citep{Daducci2015}, adapted to our substrate to estimate $D_{i,s}$, and to avoid unfair penalization of the model with parameters not relevant in our substrates.  The AMICO kernels were generated for seven radii $\tilde{R} = \{0.5, 1, 2, 3, 4, 5, 6\} \mu m$ and five diffusion coefficient $\tilde{D} = \{0.5, 1, 1.5, 2, 2.5\} \frac{\mu m^2}{ms}$, that included the ground truth values of our experiments. The soma kernels were computed for each pair $ (R_s, D_{i,s}) = (\tilde{R}, \tilde{D})$ and the extracellular kernels for each $D_{ex} = \tilde{D}$, leading to 40 kernels. As this work aimed to evaluate the impact of neglecting soma permeability on SANDI estimates, the fraction of intraneurite signal was merged with the extracellular compartment after the optimization to exclude the neurite compartment not included in our simulation substrate.

\paragraph{VERDICT}
The VERDICT (Vascular, Extracellular and Restricted Diffusion for Cytometry in Tumors) model \citep{Panagiotaki2014} is also a three-compartment model, designed for cancerous tissue, and composed of a vascular, an intracellular and an extracellular-extravascular compartment. Intracellular compartment was modeled by diffusion inside impermeable spheres of diffusivity $D_{i,s}$ \citep{Neuman1974}. Diffusion inside vascular and extra compartments were modeled as gaussian diffusion using a diffusion tensor model \citep{Basser1994}, in one principal direction $(\phi, \rho)$ with a pseudo-diffusion coefficient $P$ for the first and isotropically with diffusivity $D_{ex}$ for the second. 

In~\cite{Panagiotaki2014}, the diffusion coefficients $D_{i,s}$ and $D_{ex}$ were fixed $0.9\frac{\mu m^2}{ms}$, and the pseudo-diffusion coefficient was constrained to be larger than $3 \frac{\mu m^2}{ms}$ to reduce the parameter space.  The signal fractions, the mean cell size $R_s$ and the main orientation of the vascular compartment $(\phi, \rho)$ were estimated from the signal $S_{VERDICT}$ acquired with a gradient in direction $g$

\begin{equation}
    S_{VERDICT} = f_i S_i + f_v S_v + f_{ex} S_{ex}. 
\end{equation}

Similarly to the neurites compartment of SANDI, the volume fraction of the vascular compartment of VERDICT, which was not included in our simulation substrate, was merged with the extra-cellular space. The remaining parameters were the intracellular signal fraction  $f_i$, the mean cell $R_s$, and the intracellular $D_{i,s}$ and extracellular $D_{ex}$  diffusion coefficients. $D_{ex}$ was freed similarly to the AMICO implementation of VERDICT \citep{Bonet2019}.

\paragraph{CEXI}
 The VERDICT and SANDI models were developed to estimate soma or cell mean $R_s$ in impermeable tissues. Conversely, based on the model of exchange proposed by \cite{Karger1985}, CEXI is a two-compartment model that includes exchange between spherical cells and the extracellular space, and is the counterpart of NEXI \citep{jelescu2022} for cellular permeable tissue. The neurite compartment is replaced by a spherical compartment \citep{Neuman1974} to account for spherical cells of radius $R_s$ and intra-diffusivity $D_{i,s}$, and the extracellular diffusion is assumed Gaussian and isotropic with a diffusivity $D_{ex}$. The signal $S_{CEXI}$ is a sum of two decaying exponentials with apparent diffusion coefficient and signal fraction $(D_i', f')$ and $(D_e', 1-f')$ that are dependent on the exchange rate $r = \frac{1}{\tau_{ex}}$ between the compartments

\begin{equation}
    \fontsize{9}{12}\selectfont
    \left\{
    \begin{aligned}
    S_{CEXI}    &  = f' e^{-q^2t D_i'} + (1-f') e^{-q^2tD_e'},\\
    D_{i,e}'    & = \frac{1}{2} \left(D_i + D_{ex} + \frac{1}{q^2\tau_{ex}} \mp \left[\left[ D_{ex} - D_i + \frac{2f_i -1}{q^ 2\tau_{ex}}\right]^2 + \frac{4f_i\left(1-f_i\right)}{q^4\tau_{ex}^2}\right]^{\frac{1}{2}} \right),\\
    f'          & = \frac{1}{D_i'-D_e'} \left(f_i D_i + (1-f_i) D_{ex} - D_e'\right),\\
    D_i         & = \frac{2}{ \delta^2 D_{i, s} \left(\Delta - \frac{\delta}{3}\right)} \left[\sum_{m=1}^{\infty} \frac{\alpha_m^{-4}}{\alpha_m^2R_s^2 - 2} \left[ 2\delta - \frac{2+e^{-\alpha_m^2 D_{i,s} \left(\Delta-\delta\right)} -2e^{-\alpha_m^2D_{i,s}\delta} - 2e^{-\alpha_m^2D_{i,s}\Delta} + e^{-\alpha_m^2D_{i,s} \left(\Delta+\delta\right)}} {\alpha_m^2D_{i,s}} \right]\right],
    \end{aligned} 
    \right.
\end{equation}

where $q^2 = \frac{b}{\Delta}$ is the wavenumber of the PGSE sequence. With this formulation, CEXI has 5 parameters: the intracellular $D_{i,s}$ and extracellular $D_{ex}$ diffusion coefficients, the membrane permeability $\kappa$, the intracellular signal fraction $f_i$ and the cell radius $R_s$. Similarly to VERDICT, the parameters were fitted to the gradient encoded signals. 
\subsection{Experiments}\label{sec:experiments}
\subsubsection{Sensitivity and reliability analysis of Monte-Carlo simulations in permeable substrates}\label{sec:method_validation}

The first experiment aimed to evaluate the reliability and repeatability of the simulated signals with the substrates' biological properties described in Section~\ref{sec:method_mcdc}, \ie~ the permeability $\kappa$, the intracellular $D_{i,0}$ and the extracellular $D_{e, 0}$ diffusion coefficients. This sensitivity analysis was performed on the most challenging substrate (Table~\ref{tab:substrate}, $M_1$) to ensure accurate simulations in all substrates. The particle density  was fixed to $ \rho = 2~ m^{-3}$,  and the signals were generated with the PGSE sequence shown in Table~\ref{tab:pgse}, $E_1$. As a first validation step, the normalised mean square error ($NMSE$) of the signals was calculated with bootstrapping. 

Afterwards, the influence of the substrate properties on the time-dependency of the $ADC(t)$ and the $ADK(t)$ was investigated. The $ADC(t)$ and the $ADK(t)$ were computed from the particles' trajectories in substrates with different permeability $\kappa$ and extracellular diffusion coefficients $D_{e, 0}$. Because tracking the particles' relative position in the substrate is possible with Monte-Carlo simulations, the $ADC(t)$ and the $ADK(t)$ of the intracellular ($ADC_{in}(t)$, $ADK_{in}(t)$) and the extracellular ($ADC_{ex}(t)$, $ADK_{ex}(t)$) compartments were also calculated independently. To this end, a particle was assigned to a compartment at initialisation for the entire simulation. In the impermeable substrates, the particles in different compartments formed two independent pools. As soon as the particles diffused across the membranes, the distinction between the extracellular and intracellular compartments became ambiguous and made unclear the physical interpretation of the $ADC$ and $ADK$ of the compartment. However, their long-time limits were still used for validation of the simulator. As a last step of this time-dependency analysis, the impermeable $ADK_{ex}(t)$ was compared to the analytical solution of the diffusion kurtosis $K(t) \propto \frac{\ln(t)}{t}$ derived in the structural disorder approach \citep{novikov2014, Burcaw2015}. 

The validation experiment ended with a comparison of the $ADC(t)$ and $ADK(t)$ estimated with the signal and calculated from the propagator. Computing the $ADC(t)$ and the $ADK(t)$ from the particles' trajectories is a memory and time-consuming operation, conversely to the efficient calculation of the $ADC(t)$ and $ADK(t)$ from the DW-MRI signals. Therefore, the signal estimates were compared to the propagator ground-truth to replace the former with the latter.

\subsubsection{Comparison of compartmentalised models in permeable tissue}\label{sec:method_model}

The second experiment of this work compared the performance of the compartmentalised models VERDICT \citep{Panagiotaki2014} and SANDI \citep{Palombo2020a} designed for impermeable tissues with spherical biological structures to the compartmentalised model with exchange CEXI. Recently, the three-compartment model eSANDIX was also proposed to model exchange in gray matter \citep{Olesen2022}. This model combined the exchanging compartments of NEXI \citep{jelescu2022} and the soma compartment of SANDI. Because the numerical substrates of this work did not include a neurite compartment, this model would have been equivalent to SANDI and was therefore not tested in this work.

The results of SANDI and VERDICT were presented with different PGSE sequences in their original papers. Hence, the models were fitted to both signals generated with a model-specific PGSE sequence and with the same sequence for a baseline comparison. The SANDI specific sequence had 9 $b$ with a larger $b_{max} = \{0, 1, 2.5, 3, 4, 5.5, 7, 8.5, 10\} \frac{ms}{\mu m^2}$,  but shorter diffusion time $\Delta = \{11, 20\} ms$,  pulse length  $\delta = 3ms$  and echo time $TE = 30 ms $ (Table~\ref{tab:pgse}, SANDI). The original VERDICT PGSE sequence was downloaded from the CAMINO website \citep{Cook2005}. The echo time $TE$ of the scheme file was modified to guarantee feasible signal generation, but the original four $\Delta = \{10, 20, 30, 40 \} ms$ and  two $\delta = \{ 3 ms  $  for all , $10 ms$  for  $\Delta = \{30,40\} ms\}$ were kept (Table~\ref{tab:pgse}, VERDICT). For the baseline comparison, the PGSE sequence presented in NEXI was chosen. The DW-MRI signals were generated in $N=24$ directions for four $\Delta = \{12, 20, 30, 40 \} ms$, five $b = \{0, 1, 2.5, 4, 5 \frac{ms}{\mu m^2}\}$ and one $\delta = 4.5ms$ (Table~\ref{tab:pgse}, NEXI). 

Simulations and signals were generated with the simulator described in Section~\ref{sec:method_mcdc} and validated with the previous experiment of Section~\ref{sec:method_validation}. For all simulations, the particle density $\rho$ and the intracellular diffusion coefficient $D_{i, 0}$ were fixed to $0.5 m^{-3}$ and $2 \frac{\mu m^2}{s}$ respectively. The permeability $\kappa$ was set to $0, 1, 25$ or $50 \frac{\mu m}{s}$ , and the extracellular diffusion coefficient $D_{e, 0}$ to $0.5, 1 $ or $2 \frac{\mu m^2}{s}$. All $(\kappa, D_{e, 0})$ combinations were simulated (Table~\ref{tab:experiment}, $E_2$). Additionally, 30 corrupted signals with Rician noise of $SNR=\{30; 80\}$ were generated using the DIPY software library \citep{Garyfallidis2014}.

Models were fitted to the signals for each diffusion time $\Delta$ independently and all $\Delta$ simultaneously. SANDI was fitted with the adapted AMICO implementation, while VERDICT and CEXI were fitted to the signals following the guidelines described in~\cite{Alexander2008} with the constrained least square implementation of the python optimisation library LEVMAR \citep{lourakis04LM}. The signal fractions were constrained to sum up to 1 and boundaries were imposed on the value of the parameters to avoid unrealistic estimations: $\overline{R} \in \left[0.1, 20\right] \mu m , \overline{ICVF} \in \left[0.1, 0.9\right], (\overline{D}_i ,\overline{D}_e)  \in \left[0.01, 3\right] \frac{\mu m^2}{ms}$  for both models, and $\kappa  \in \left[0, \infty\right[ \frac{\mu m}{s}, \lambda \left[1, \infty\right[ $ for CEXI. Ten optimizations with random initialisation were performed and the selected estimation was the least costly.

The models were compared based on their estimates of the mean cell radius $\overline{R}$, the intracellular volume fraction $\overline{ICVF}$ and the extracellular diffusion coefficient $\overline{D}_e$. The ground truth values of the intracellular volume fraction $ICVF$ and the volume-weighted mean cell radius $R$ were calculated from the substrates, (Table~\ref{tab:substrate}), while the apparent extracellular diffusion $D_e$ was calculated with the propagator for each experiment. Due to the obstacles encountered by the particles, this apparent extracellular diffusion coefficient might differ significantly from $D_{e, 0}$ with the diffusion time. Following to the time-dependency analysis of Section~\ref{sec:method_validation}, the $D_e$ was calculated from the propagator of the extracellular compartment at the longest diffusion time. Because the models must disentangle the compartments even in the presence of exchange, the $ADC_{ex}(t)$ of the impermeable substrate was considered the ground-truth, \ie~ $D_e = ADC_{ex}(t=\Delta_{max}, \kappa=0)$.

\section{Results}\label{sec::results}
\subsection{Influence of the substrate properties and the simulation parameters on the Monte-Carlo simulated DW-MRI signals}

\begin{figure}[h!]
\begin{center}
\includegraphics[width=\textwidth]{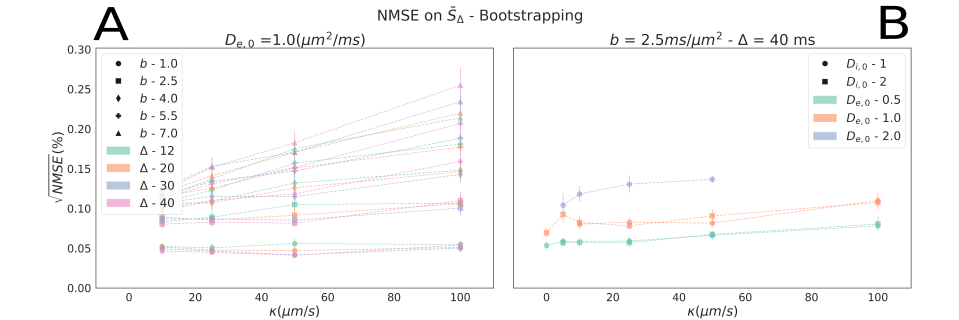}
\end{center}
\caption{\textbf{Normalised mean squared error on the signal.} Normalised mean squared error $(NMSE)$ on the signal with the permeability $\kappa$ from bootstrapping for all pairs of $b$ and diffusion time $\Delta$ (A) and for all diffusion coefficients $\Delta$ (B). (A) The $NMSE$ is shown for $D_{e, 0}=1\frac{\mu m^2}{ms}$ and $D_{i, 0}=2\frac{\mu m^2}{ms}$. The error increased with $\kappa$ for all $b$ - $\Delta$ pairs, and its range for different $\Delta$ broadened as $b$ increased. (B) The $NMSE$ is shown for the pair ($b=2.5 \frac{ms}{\mu m^2}, \Delta=40ms)$, for different $D_{e, 0}$ (colour) and  $D_{i, 0})$ (symbol). The error increased with $D_{e,0}$ but seemed independent on $D_{i,0}$. With the chosen simulation parameters, the $\sqrt{NMSE}$ never exceeded $0.3\%$. }\label{fig:signal_error}
\end{figure}

Fig. \ref{fig:signal_error} shows the evolution of the bootstrapped $NMSE$ on the DW-MRI signal with the permeability $\kappa$ for all pairs ($b$, diffusion time $\Delta$) (Fig.~\ref{fig:signal_error}A) and all combinations of the diffusion coefficients $(D_{i, 0}, D_{e,0})$ (Fig.~\ref{fig:signal_error}B). For all permeability $\kappa$ and diffusion time $\Delta$, the $NMSE$ increased with $b$ (Fig.~\ref{fig:signal_error}A) and,  at large $b$ ($b > 4 \frac{ms}{\mu m^2}$), the error became also dependent on $\Delta$ and $\kappa$. 

When the $b$ and the $\Delta$ of the PGSE sequence were fixed to $(b=2.5 \frac{ms}{\mu m^2}, \Delta =40ms)$ (Fig. \ref{fig:signal_error}B), the diffusion coefficients had distinct effect on the error. Indeed, the $NMSE$ was dependent on $D_{e,0}$ (colours) while it seemed independent on $D_{i, 0}$ (markers). Overall, the maximal error was reached with the largest $b$, the longest $\Delta$ and the most permeable membrane but remained under $0.3\%$ with the chosen simulation parameters. 

The results of the time-dependency analysis of the $ADC(t)$s and the $ADK(t)$s are shown in Fig.~\ref{fig:signal_ADC_ADK_in_ex}. In the impermeable substrate (green lines), the extracellular $ADC_{ex}(t)$ (square) converged quickly to its long-time limit $1.25 \frac{\mu m^2}{ms}$, while the intracellular $ADC_{in}$ (circle) decreased towards zero (Fig.~\ref{fig:signal_ADC_ADK_in_ex}A), with a decay rate similar to the $ADC(t)$ convergence rate calculated with all particles (diamond). As the permeability increased, the $ADC_{in}(t)$ and the $ADC_{ex}(t)$ converged faster to the long-time limit of the $ADC(t)$.

Similarly, the time-dependency of the intracellular $ADK_{in}(t)$ and the extracellular $ADK_{ex}(t)$ had a different trend in impermeable and permeable substrates (Fig.~\ref{fig:signal_ADC_ADK_in_ex}B). The $ADK_{in}(t)$ and the $ADK_{ex}(t)$ stabilized around 2 and 0 in the former, whereas both peaked before decreasing in the latter. In Fig.~\ref{fig:signal_ADC_ADK_in_ex}C, this $ADK_{ex}(t)$ in impermeable substrates is plotted against $\frac{\ln(t)}{t}$ for different extracellular diffusion coefficient $D_{e, 0}$ (colours). After a transition phase at the beginning of the simulation, the $ADK_{ex}(t)$ evolved linearly with $\frac{\log(t)}{t}$ for all $D_{e, 0}$ matching the power-law of the structural-disorder theory.

\begin{figure}[h!]
\begin{center}
\includegraphics[width=\textwidth]{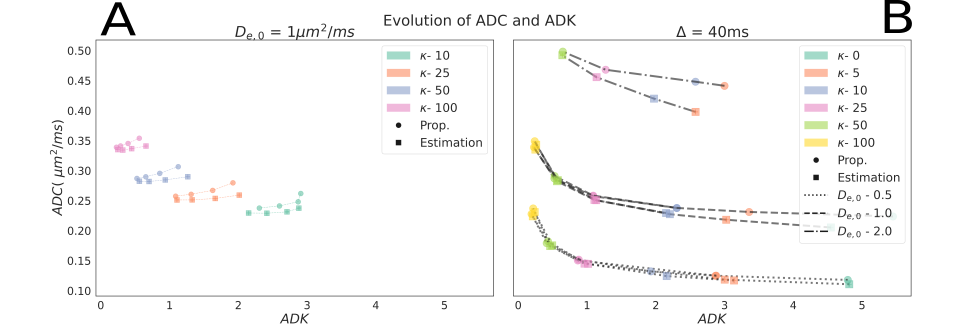}
\end{center}
\caption{\textbf{Signal and propagator $\mathbf{ADC(t)}$ and $\mathbf{ADK(t)}$.} Comparison of the $ADC(t)$ and the $ADK(t)$ estimated from the signal (square) and the propagator (circle) for different permeability $\kappa$ (colour) with the diffusion time $\Delta$(A) and the diffusion coefficients ($D_{i, 0}, D_{e,0}$) (B). The symbols are located at the mean signal-$ADC$ and -$ADK$ of the bootstrapped signals. (A) The $ADC(t)$s and the $ADK(t)$s  are shown for the simulations with an extracellular diffusion coefficient $D_{e, 0} = 1 \frac{\mu m^2}{ms}$. The symbols on the same line had the same $\kappa$ but a different $\Delta$. The $ADC(t)$ and the $ADK(t)$ estimated with the signal and the propagator were close for all $\kappa$. The $ADC(t)$s increased with $\kappa$ while $ADK$s decreased. (B) The $ADC$s and the $ADK$s are plotted for one diffusion time $\Delta = 40 ms$. For $D_{e,0}=\{1, 2\}\frac{\mu m^2}{ms}$ , two lines with the same style are plotted corresponding to $D_{i,0}=\{1, 2\}\frac{\mu m^2}{ms}$. $D_{i,0}$ is shown without legend because the influence of the intracellular diffusion coefficient was negligible in comparison to the $D_{e,0}$ or $\kappa$. }\label{fig:signal_ADC_ADK}
\end{figure}

Fig.~\ref{fig:signal_ADC_ADK} shows the $ADC(t)$ and $ADK(t)$ of the full signal estimated both from the signal (square) and the propagator (circle) for different $\kappa$. The symbols on the same line plot in Fig.~\ref{fig:signal_ADC_ADK}A are the $ADC$-$ADK$ estimated for the same experiment at different $\Delta$. For all experiments, the $ADC$ calculated with the propagator was slightly greater than the $ADC$ calculated from the signal, but both had similar trends. In the range of $\Delta$, the $ADC$ and $ADK$ from the propagator decreased at most by $0.02 \frac{\mu m^2}{ms} $ and $0.6$ respectively, at the smallest permeability $\kappa = 10 \frac{\mu m}{s}$. At the largest $\kappa=100 \frac{\mu m}{s}$, the $ADC$ decreased by $0.015 \frac{\mu m^2}{ms}$, and the $ADK$ $0.33$, suggesting that more permeable substrates reached faster an equilibrium.

Fig.~\ref{fig:signal_ADC_ADK}B  shows that both $\kappa$ (colours) and $D_{e, 0}$(lines) increased the $ADC$ and different experiments led to the same $ADC$ for different pair ($\kappa$, $D_{e, 0}$). Similarly, the experiments with the same $\kappa$ had an $ADK$ in the same range for all  $D_{e,0}$. Finally, the $ADC$ and $ADK$ of experiments that differ by $D_{i,0}$ only were indistinguishable.

\subsection{Sensitivity of compartmentalised models to the substrate properties}\label{sec:results_model}

SANDI(A), VERDICT(B) and CEXI(C) were fitted to the noiseless signals using all diffusion times $\Delta$ in the substrates with one (Fig.~\ref{fig:model_single_sphere_allmethods}) or two (Fig.~\ref{fig:model_mul_sphere_allmethods}) populations of spheres. For each model, the subplots show the estimates (top) and the Mean Absolute Error (MAE, bottom) on the mean radius $\overline{R}$ (left column), the extracellular diffusion coefficient $\overline{D}_e$ (centre column) and the intracellular volume fraction $\overline{ICVF}$ (right column) with respect to the exchange time $\tau_{ex}$. In all subplots, the ground-truth permeability $\kappa$ is encoded with the colours and the effective extracellular diffusion coefficient $D_e$ with the symbols. The vertical dashed lines separate the results generated with the same substrates, \ie~ $(S_1, S_2, S_3, S_4)$ with $R = \{2, 3, 4, 5\} \mu m$ in Fig.~\ref{fig:model_single_sphere_allmethods} and  $(M_1, M_2, M_3)$ with $(R_1, R_2) = \{(1,3) ; (1,5); (3,5)\}  \mu m$ in Fig.~\ref{fig:model_mul_sphere_allmethods}). In the supplementary materials, the estimations and the MAE fitted with all $\Delta$ independently and on noisy data are available, as well as the results of SANDI and VERDICT with their respective PGSE sequences.

\subsubsection{Mean cell size estimation}
\begin{figure}[h!]
\begin{center}
\includegraphics[width=\textwidth]{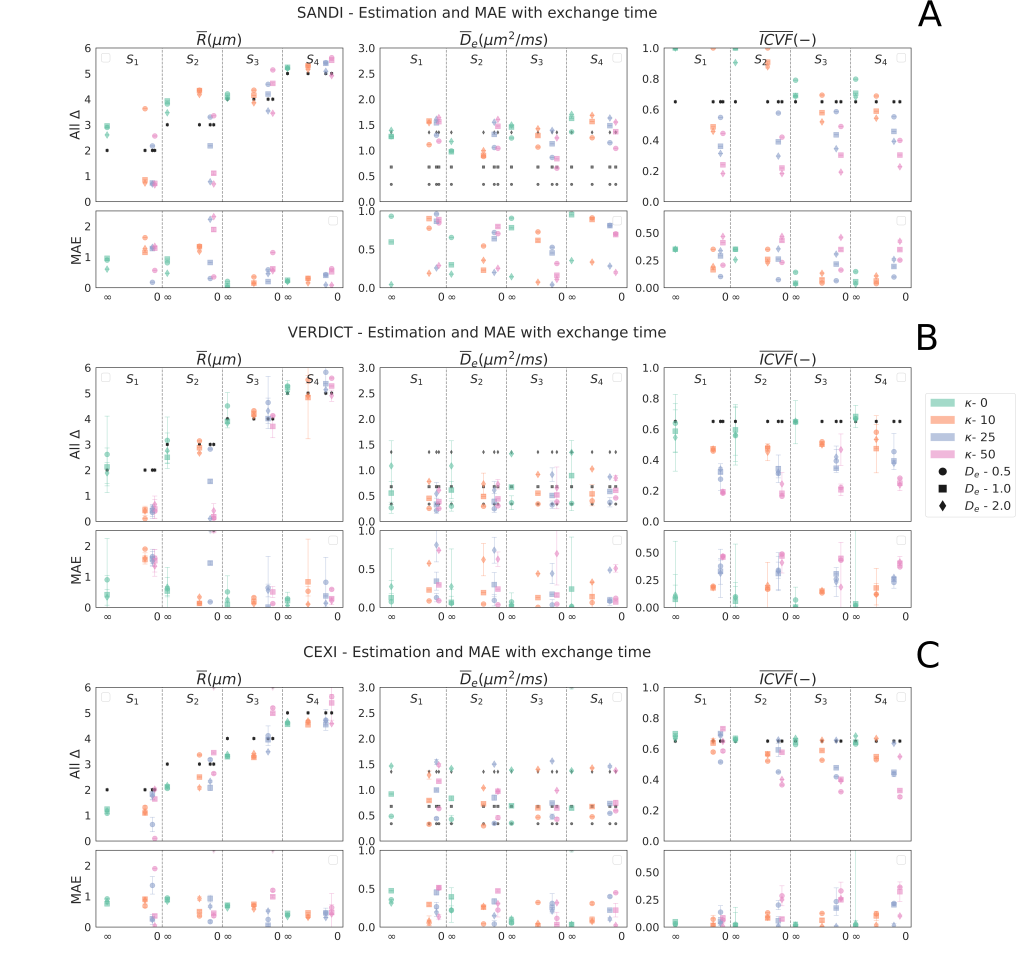}
\end{center}
\caption{ \textbf{Results of compartmentalized models on the single-$\mathbf{R}$ substrates.} Estimates and MAEs of the compartmentalized models SANDI (Row A), VERDICT(Row B) and CEXI (Row C) of the single-$R_s$ substrate properties, from all diffusion times $\Delta$. For each subplot, vertical lines separate the substrates $S_{i}$ of Table~\ref{tab:substrate} with a different mean sphere radius $R_S$. Between these lines, the location of the experiments on the x-axis are determined by the exchange time $\tau_{ex}$. Experiments with the same permeability $\kappa$ are shown in the same colour, and the symbols encodes the effective extracellular diffusion coefficient $D_e$. The symbols are located at the mean estimates of the experiments and the bars shows the variance across estimates. (Left column) The mean volume-weighted radius $R$ was well estimated by SANDI and VERDICT in substrates with $R > 4\mu m$ or in impermeable substrates. CEXI estimates were nearly independent of the permeability $\kappa$ and improved with larger $R$. (Centre column) The MAEs on the effective extracellular diffusion coefficient $D_e$ were large for all methods. SANDI and VERDICT poorly captured variations of $D_e$ in the case of permeable cells and the $\overline{D}_e$ were overall independent on the underlying ground truth values. CEXI $\overline{D}_e$ was sensitive to $D_e$ for all $R$ and $\kappa$. (Right) The MAEs of $\overline{ICVF}$ decreased with $\kappa$ for all models. The SANDI and VERDICT MAE's were fairly independent of $R$, while the CEXI error increased with $R$.}\label{fig:model_single_sphere_allmethods}
\end{figure}

\begin{figure}[h!]
\begin{center}
\includegraphics[width=\textwidth]{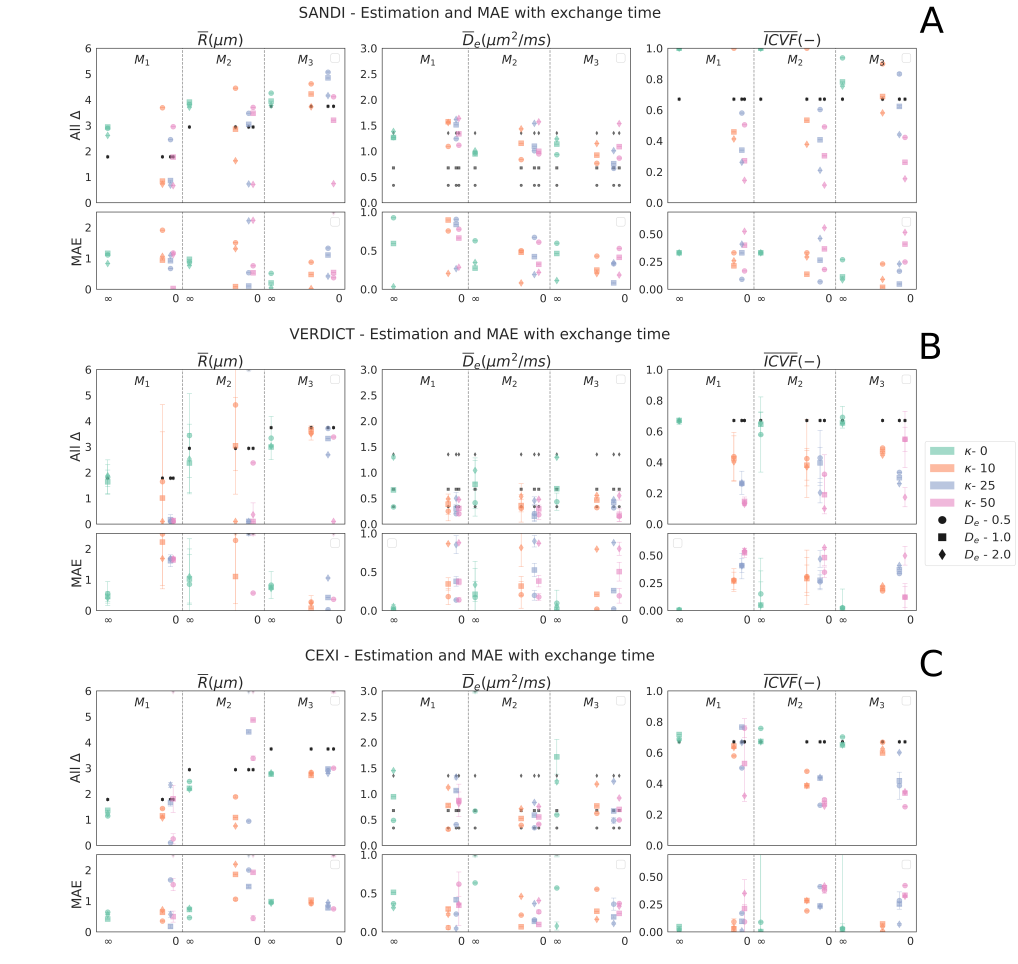}
\end{center}
\caption{\textbf{Results of compartmentalized models on the multiple-$\mathbf{R}$'s substrates.}Estimates and MAEs of the compartmentalized models SANDI (Row A), VERDICT(Row B) and CEXI (Row C) of the multiple-$R_s$ substrate properties, from all diffusion times $\Delta$. Plots are similar to Fig.\ref{fig:model_single_sphere_allmethods} but for the multiple-$R_s$ substrates of Table~\ref{tab:substrate}. (A) $R$ and $ICVF$ were well estimated in the impermeable substrates only. (B) The MAE of $\overline{R}$ was acceptable for the largest $R$ only, but the MAE on the $ICVF$ was under $0.1$ in all impermeable substrates. (C) The MAEs on $R$ were in the same range for different $\kappa$ in the substrates $M_1$ and $M_3$ with a small difference between the radii. Again, the $ICVF$ was better estimated in substrates with a small gap between the spheres' radii and $\kappa < 10 \frac{\mu m}{s}$.
 }\label{fig:model_mul_sphere_allmethods}
\end{figure}

The MAE of the radius estimation $\overline{R}$ increased as the cell size $R$ decreased and the permeability $\kappa$ increased, for all models (Fig.~\ref{fig:model_single_sphere_allmethods}, left column). Consequently, the maximal substrate permeability $\kappa$ to ensure a MAE below a targeted threshold was larger in bigger spheres. For example, the SANDI MAE (Fig.~\ref{fig:model_single_sphere_allmethods}A) remained under $0.6\mu m$ in the substrates $S_2, S_3$ and $S_4$ if $\kappa$ was below $10\frac{\mu m}{s}, 25 \frac{\mu m}{s}$ and $50 \frac{\mu m}{s}$ respectively. For all multiple-$R_s$ substrates, SANDI $\overline{R}$ was close to $R$ in impermeable experiments only (Fig.~\ref{fig:model_mul_sphere_allmethods}A). 

With VERDICT, the MAE on $R$ in substrates $S_3$ and $S_4$ was stable with $\kappa$ and remained under $0.6\mu m$, while in substrates $S_1$ and $S_2$ either the variance of $\overline{R}$ was large, or the mean $\overline{R}$ was not accurate (Fig.\ref{fig:model_single_sphere_allmethods}B). 
 In the multiple-$R_s$ substrates, the MAE of $\overline{R}$ remained under $1 \mu m$ with impermeable membrane, or in the substrate $M_3$ with the biggest spheres (Fig.~\ref{fig:model_mul_sphere_allmethods}B).

Conversely to SANDI and VERDICT, the CEXI MAE of $\overline{R}$ in single-$R_s$ substrates was nearly independent on $\kappa$ for all $R$ (Fig.\ref{fig:model_single_sphere_allmethods}C). The substrates $S_1$ with the smallest $R$ and $S_4$ with the largest $R$ had a maximal MAE of $1\mu m$ and $0.5 \mu m$ for most of the exchange times, respectively. In multiple-$R_s$ substrates, CEXI underestimated the mean $R$ in almost all cases (Fig.\ref{fig:model_mul_sphere_allmethods}C). The MAE remained smaller than $ 1\mu m$ in the substrate $M_3$ for all experiments, and in the substrate $M_1$ for $\kappa < 25 \frac{\mu m}{s}$. 

Both for SANDI and VERDICT, the $\overline{R}$ MAE increased with a slower $D_e$ in substrates $S_3$ and $S_4$ with $R > 3\mu m$ (Fig.\ref{fig:model_single_sphere_allmethods} A-B), while the CEXI $\overline{R}$ seemed independent on $D_e$ (Fig.\ref{fig:model_single_sphere_allmethods}C, Fig.\ref{fig:model_mul_sphere_allmethods}C).

\subsubsection{Extracellular diffusion coefficient estimation}
For all experiments and models, the permeability $\kappa$ didn't influence the extracellular diffusion coefficient estimation $\overline{D_e}$ (Fig.\ref{fig:model_single_sphere_allmethods}, \ref{fig:model_mul_sphere_allmethods}).  On the one hand, VERDICT and SANDI poorly captured variations of $D_e$ in the case of permeable cells. SANDI $\overline{D_e}$ over-estimated $D_e$ in most cases, and was overall independent on the underlying ground truth values. VERDICT estimated $D_e$ accurately in the case of $\kappa=0$, but underestimated $D_e$ in permeable substrate. On the other hand, CEXI showed good performance in estimating $D_e$, for all effective $D_e$ and ground-truth $\kappa$ values (Fig.\ref{fig:model_single_sphere_allmethods}C) 

\subsubsection{Intracellular volume fraction estimation}
The MAEs of $\overline{ICVF}$ increased with $\kappa$ for all models both in single-$R_s$ (Fig.\ref{fig:model_single_sphere_allmethods}) and multiple-$R_s$ substrates (Fig.\ref{fig:model_mul_sphere_allmethods}). The error was fairly independent of cell size for SANDI and VERDICT, while in CEXI the error increased with $R$ (Fig.\ref{fig:model_single_sphere_allmethods}). In the substrates $M_1$ and $M_3$, CEXI estimated accurately the $ICVF$ in the low permeability regime (Fig.\ref{fig:model_mul_sphere_allmethods}). This is consistent with the Kärger model of exchange holding in the case of well-mixed compartments with barrier-limited exchange and thus breaking for larger cells.
    
\subsection{Permeability estimation with CEXI}\label{sec:results_perm}
\begin{figure}[h!]
\begin{center}
\includegraphics[width=\textwidth]{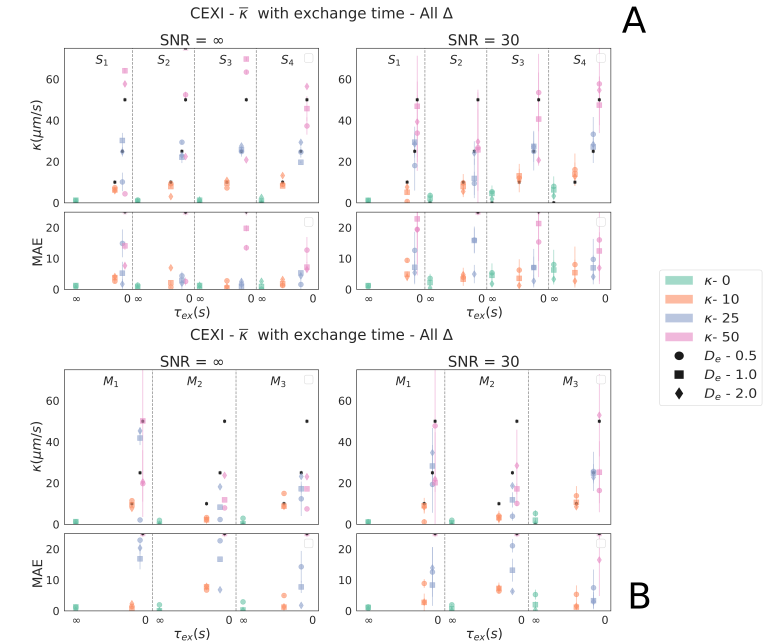}
\end{center}
\caption{ \textbf{CEXI permeability estimates.} Estimate and MAE of the permeability $\kappa$ with CEXI in single-$R_s$ (Row A) and multiple-$R_s$(Row B) substrates (Table~\ref{tab:substrate}) from all diffusion times $\Delta$ using noise-free signals (left column) or noisy data (right column). (Left) CEXI estimated well the moderate permeability $\kappa < 25 \frac{\mu m}{s}$ for all the single-$R_s$ substrates (A) and the multiple-$R_s$ substrates $M_1$ and $M_3$ with small difference between the sphere radii (B). (Right) With noisy data, the variance increased, and the MAE increased at large $\kappa > 25 \frac{\mu m}{s}$.
 }\label{fig:cexi_tex}
\end{figure}

CEXI is the only compartmentalised model that estimated the exchange time $\tau_{ex}$ and, therefore, the permeability $\kappa$ between intra and extracellular compartments. Fig.~\ref{fig:cexi_tex} shows the permeability estimate $\overline{\kappa}$ and its associated MAE with all diffusion times $\Delta$ in single-$R_s$ (Fig.~\ref{fig:cexi_tex}A) and multiple-$R_s$ substrates (Fig.~\ref{fig:cexi_tex}B). Overall, in the single-$R_s$ substrates, the permeability estimate $\overline{\kappa}$ correlated well with the true permeability $\kappa$, although the error increased with $\kappa$. The error was also larger in substrates with a slower $D_e$, while the cell radius $R$ had a limited effect. In multiple-$R_s$ substrates, CEXI tended to underestimate the permeability $\kappa$. 

For a fixed $\kappa$, the error on $\overline{\kappa}$ was similar for the different mean cell sizes in single-$R_s$ substrates (Fig.~\ref{fig:cexi_tex}A). For all $R$, although the MAE of $\overline{\kappa}$ increased with $\kappa$, the relative error decreased with $\kappa$ and $R$. For example, the maximal relative error on $\kappa=50\frac{\mu m}{s}$ went from $90 \%$ in the substrate $S_1$ with the smallest sphere down to $25\%$ in $S_4$ with the biggest spheres. 

The estimates with a single $\Delta$ are available in the supplementary material. The model was not sensitive to the variation of $\kappa$ when estimated with a single $\Delta$.


\section{Discussion}\label{sec::discussion}
\subsection{Validation of the Monte-Carlo simulations in permeable tissue}\label{sec:discussion_signal}

This work investigated through realistic Monte-Carlo simulations the effect of membrane permeability $\kappa$ on the estimation of microstructure models parameters. To perform simulations in permeable substrates with multiple diffusivities, the open-source MC-DC diffusion simulator \citep{Patino2020} was extended following previous studies \citep{Szafer1995, Lee2021}. The reliability of the signal generation was validated by a small $NMSE$ (Fig.~\ref{fig:signal_error}) and a time-dependency of the $ADC(t)$ and the $ADK(t)$ (Fig.~\ref{fig:signal_ADC_ADK_in_ex}, \ref{fig:signal_ADC_ADK}) consistent with the power-law predicted by structural disorder, in the long time limit. \citep{novikov2014, Burcaw2015}.  

\subsubsection{Repeatability of the signal generation}

The permeability $\kappa$ was identified as a first important substrate parameter to guarantee the repeatability of the simulation and a small $NMSE$, especially at a high $b$-value (Fig.~\ref{fig:signal_error}A). The increase in the $NMSE$ with $\kappa$ was faster at high $b$, and the $NMSE$ range for different diffusion times $\Delta$ broadened as $b$ increased. This suggests that the particle density required for the simulations is more dependent on $b$ and $\kappa$ than $\Delta$. In parallel, the extracellular diffusion coefficient $D_{e, 0}$ was shown to play a major role in signal generation, which confirms that both intracellular and extracellular signals must be simulated simultaneously (Fig.~\ref{fig:signal_error}B). Despite the increase in the $NMSE$ with $\kappa$ and $b$, the small error  $(\sqrt{NMSE} < 0.3\%)$ of the signals validated the choice of the simulation parameters for the model comparison experiment (Section \ref{sec:method_model}).

\subsubsection{Is the simulated signal consistent?}

\begin{figure}[h!]
\begin{center}
\includegraphics[width=\textwidth]{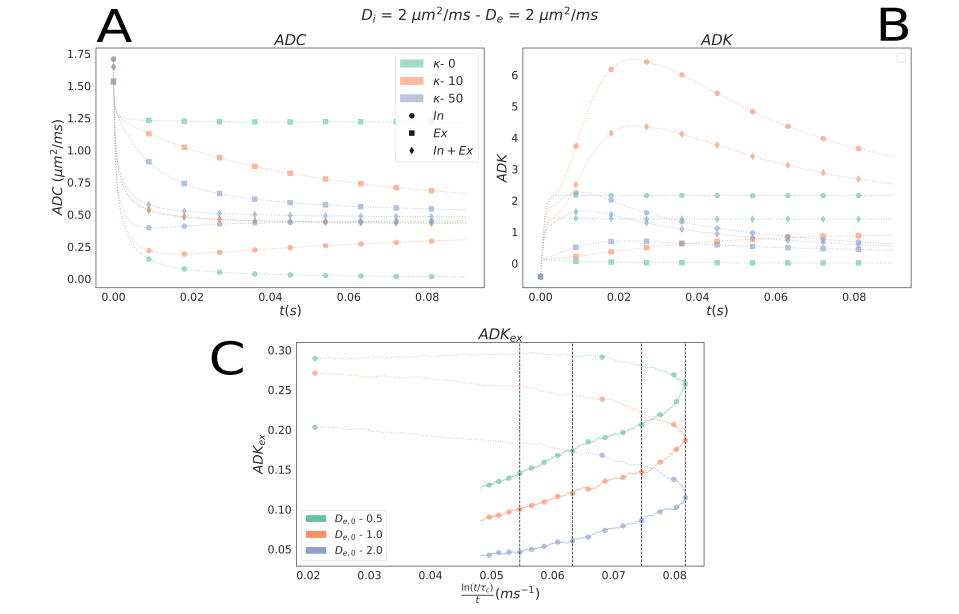}
\end{center}
\caption{\textbf{ $\mathbf{ADC(t)}$ and $\mathbf{ADK(t)}$ time-dependency.} Time-dependency of the $ADC(t)$ (A) and the $ADK(t)$ (B) of the intracellular compartment (circle), the extracellular compartment  (square) and both compartments (diamond) of the substrate $M_2$ (Table~\ref{tab:substrate}) estimated from the propagator, for different permeability $\kappa$ (colour). (A) In the impermeable substrate, $ADC_{in}(t)$ converged to $0\frac{\mu m^2}{ms}$ and $ADC_{ex}(t)$ reached its long-time limit in less than $20 ms$. As the permeability increased, the compartments mixing time shortened, and the $ADC_{in}(t)$ and $ADC_{ex}(t)$ converged faster to the long-time limit of the signal-$ADC(t)$. (B) In impermeable substrate, the $ADK_{in}(t)$ and the $ADK_{ex}(t)$ converged also to their long-time limits in less than $20 ms$. In permeable substrates, the signal-$ADK(t)$ reached a peak before decreasing. As the permeability increase, the peaking time was shifted to shorter diffusion time and the maximal $ADK(t)$ was reduced. (C) The time-dependency of the extracellular $ADK_{ex}(t)$ of the impermeable substrate is plotted against $\ln{(t)}/t$. After a transition time around $10ms$, the time-dependency of the $ADK_{ex}(t)$ became linear with $\ln{(t)}/t$ for all extracellular diffusion coefficients $D_{e,0}$ (colour), matching the power-law of the structural-disorder theory (~\cite{novikov2014, Burcaw2015}).
}\label{fig:signal_ADC_ADK_in_ex}
\end{figure}

After validating the repeatability of the simulations, the time-dependency of the $ADC(t)$ and the $ADK(t)$ of the intracellular and the extracellular spaces supported the validity of the simulations (Fig.~\ref{fig:signal_ADC_ADK_in_ex}). In the impermeable substrates, the $ADC_{in}(t)$ of the particles confined inside the intracellular space decreased towards $0 \frac{\mu m^2}{ms}$ (Fig.~\ref{fig:signal_ADC_ADK_in_ex}A). Simultaneously, the $ADC_{ex}(t)$ converged quickly to its long-time limit linearly with $\ln(t)/t$. (Fig.~\ref{fig:signal_ADC_ADK_in_ex}A, C). In the mesoscopic structural disorder theory, this time-dependency in the long-time limit characterizes medium with a short-range disorder in 2 dimensions \citep{Burcaw2015} or an extended-disorder in 3 dimensions \citep{novikov2014}. Future work should discuss how permeability affects the approaching regime of the long-time limit, and if this approach could discriminate healthy tissue from tumors. 

In permeable substrates, the physical interpretation of the $ADC_{in}(t)$ and the $ADC_{ex}(t)$ became ambiguous, as exemplified by the increase of the $ADC_{in}(t)$ with the diffusion time (Fig.~\ref{fig:signal_ADC_ADK_in_ex}A). However, their long-time limits still informed about the validity of the simulations because the $ADC_{in}(t)$ and the $ADC_{ex}(t)$ were expected to converge to the same value. Indeed, the diffusion properties of a particle changed when crossing the membrane. After a diffusion time sufficiently long, many particles had crossed the membranes and, therefore, the compartments appeared mixed and were no more distinguishable. Hence, the $ADC_{in}(t)$, the $ADC_{ex}(t)$ and the $ADC(t)$ must have converged to the same long-time limit. At faster permeability, the mixing rate of the compartments increased, and the $ADC(t)$s converged faster. 

In the same way, the time dependency of the $ADK(t)$ was also very dependent on permeability (Fig.~\ref{fig:signal_ADC_ADK_in_ex}B). As the permeability increase, the peaking time of the $ADK(t)$ was shifted to shorter diffusion time. These observations were coherent with previous work \citep{Aggarwal2020, Zhang2021} and provided a way to choose the parameters of the PGSE sequence to become sensitive to targeted microstructure. For example, the decaying exponential of the Kärger model \citep{Karger1985} required the kurtosis to be in the decreasing phase. To be in this regime, the DW-MRI signals of microstructure with a long exchange time due to slow permeability or big structure should be acquired with longer diffusion time. 

Apart from the Monte-Carlo simulations, the ground truth diffusion propagator is not accessible, and the $ADC(t)$ and $ADK(t)$ are computed from the signal. Therefore, the $ADC(t)$ and $ADK(t)$ calculated from the signals were compared to this ground-truth. Ultimately, the estimates of both approaches were close, and they were affected by the substrate properties in a similar way: the $ADC(t)$ increased and the $ADK(t)$ decreased with the permeability $\kappa$, respectively (Fig.~\ref{fig:signal_ADC_ADK}). The gap between the $ADC(t)$ and the $ADK(t)$ at the shortest and the longest diffusion times were bigger at short than at fast permeability $\kappa$  (Fig.~\ref{fig:signal_ADC_ADK}A).  Coupled with the time-dependency of the $ADC(t)$ and $ADK(t)$ in each compartment (Fig.~\ref{fig:signal_ADC_ADK_in_ex}A, B), these observations supported that diffusion reached the long-time limit faster in more permeable tissue.

\subsubsection{Degeneracy of substrate properties estimation in permeable tissue}

In a realistic substrate, the time-dependent $ADC(t)$ and $ADK(t)$ are impacted by the diffusivities and characteristic lengths of each compartment as well as by the exchange, rendering the estimation of all model parameters from these signal representations challenging. Notably, different pairs $(D_{e,0}, \kappa )$ were shown to yield the same $ADC(t)$ (Fig.~\ref{fig:signal_ADC_ADK}B). Experiments with a large permeability $\kappa$ and a small extracellular diffusion coefficient $D_{e,0}$ had an $ADC(t)$ similar to experiments with a larger $D_{e,0}$ but a smaller $\kappa$. For example, the pairs $(D_{e,0}, \kappa) = (0.5~\frac{\mu m^2}{s}, 100~\frac{\mu m}{s})$ and  $(D_{e,0}, \kappa) = (1~\frac{\mu m^2}{ms}, 10~\frac{\mu m}{s})$ had an $ADC(t)$ around $ 0.22~\frac{\mu m^2}{ms}$. If the extracellular diffusion coefficient $D_{e,0}$ of the tissue was known, the $ADK(t)$ could disentangled this multiplicity. In the previous example, the $ADK(t)$s were equal to $0.3$ and $2.3$ respectively. However, the $ADK(t)$ suffered also of degeneracy as substrates with the same permeability $\kappa$ but different $D_{e,0}$ had an $ADK(t)$ in the same range. This sensitivity and specificity analysis also highlighted the weak dependency of the $ADC(t)$ and the $ADK(t)$ with respect to the $D_{i,0}$ despite an acceptable $ICVF=0.65$ (Fig.~\ref{fig:signal_ADC_ADK}B), confirming the conclusions of \cite{Li2017}. This observation might explain why the estimation of the $D_{i,0}$ by compartmentalised models has been a challenging task so far \citep{jelescu2022, Karunanithy2019, Novikov2018, Palombo2020a}. Future work should confirm this conclusion in substrates with bigger cell size $R_s$, where the contribution of the intracellular signal is expected to differ from this work.

\subsection{Performance of the compartmentalized models in permeable substrates of tumors}\label{sec:discussion_model}

In the second experiment, the compartmentalized models SANDI \citep{Palombo2020a} and VERDICT \citep{Panagiotaki2014} developed for impermeable tissue with spherical microstructure were compared to CEXI, a model including exchange. From the first experiment (Section \ref{sec:discussion_signal}), the mean volume-weighted radius $R$, the effective extracellular diffusion coefficient $D_e$ and the permeability $\kappa$ were pointed out as the substrates properties with the biggest influence on the signal. Therefore, this experiment focused on their effect on the estimation of the microstructure properties. 

Previous studies demonstrated that estimation of the substrates properties with compartmentalized models is a challenging ill-posed problem. The low specificity of these models to the diffusion coefficient was highlighted in WM \citep{Jelescu2016, Li2017} and, more recently, in gray matter \citep{Palombo2020a, jelescu2022, Olesen2022}. Additionally, the size of the axons in WM \citep{Burcaw2015} or the cells in gray matter \citep{Afzali2021, Palombo2021, Olesen2022} was shown to be overestimated. Starting from this known initial state in impermeable tissue, this experiment showed the impact of permeability on the model estimates of the mean cell radius $\overline{R}$, the extracellular diffusion coefficient  $\overline{D}_e$ and the intracellular volume fraction $\overline{ICVF}$. The evolution of these estimates highlighted the distinct behavior of CEXI in comparison to SANDI and VERDICT in reaction to variations of the microstructure properties.

\subsubsection{Stability of the estimates with an increasing permeability}

VERDICT and SANDI estimates had similar trends with respect to the permeability $\kappa$ despite different performances. In impermeable substrate, the volume-weighted mean cell size $R$ and the intracellular volume fraction $ICVF$ were overestimated and underestimated as expected \citep{Afzali2021, Olesen2022}(Fig.~\ref{fig:model_single_sphere_allmethods}). As the permeability increased, the bias on the parameter estimation in impermeable substrate was amplified. Consequently, the error of $\overline{ICVF}$ and $\overline{R}$ increased with the permeability $\kappa$. This deterioration of the estimates was coherent with the recent observation that exchange dominated the signal in gray matter at long diffusion time ($\Delta>20ms$) \citep{Olesen2022}, comparable to the diffusion time investigated here. The opposite evolution of $\overline{ICVF}$ and $\overline{R}$ indicated how the models developed for impermeable tissues compensated for water exchange. Because the distance crossed by particles in impermeable cells was limited by the cell size, the increase in the $ADC$ due to permeability was compensated by either decreasing the proportion of the intracellular signal via a smaller $\overline{ICVF}$, or increasing the maximal distance via a larger $\overline{R}$. 
 
With CEXI, this effect was attenuated thanks to the exchange time $\overline{\tau}_{ex}$ capturing most of the exchange effect. The CEXI $\overline{ICVF}$ decreased with an increasing permeability $\kappa$  as in neurites \citep{jelescu2022}, and $\overline{R}$ overestimated the volume-weighted mean cell size $R$ in the substrates with the fastest permeability $\kappa$ only. At moderate permeability $(\kappa < 25 \frac{\mu m}{s})$, CEXI disentangled the effect of exchange and restriction from the DW-MRI signals, providing more stable estimates of  $\overline{ICVF}$ and $\overline{R}$ with the cell size and the permeability than VERDICT and SANDI.

\subsubsection{Sensitivity of the estimates to the extracellular diffusion coefficient}

The diffusion coefficients are arguably the most difficult parameters to estimate due to degeneracy of the solution \citep{Jelescu2016, Novikov2018} and the low sensitivity of the models to the intracellular diffusion coefficient \citep{Li2017}. In WM, the intracellular diffusion coefficient was often considered faster than the extracellular diffusion coefficient \citep{Kunz2018, Dhital2019, Olesen2021}, but recent studies in gray matter suggested contradictory conclusions on which compartment had the fastest diffusivity ($D_e > D_i$ in \citep{Olesen2022} or $D_e < D_i$ in  \citep{jelescu2022}). Interestingly, VERDICT and SANDI estimates in this experiment reflected this uncertainty. Indeed, both models were nearly insensitive to the extracellular diffusion coefficient $D_e$, but SANDI estimated a $\overline{D_e}$ larger than $D_{i,0}$ (Fig.~\ref{fig:model_single_sphere_allmethods} A, center column) while VERDICT $\overline{D_e}$ was smaller (Fig.~\ref{fig:model_single_sphere_allmethods} B, center column). In the big cells ($S_3$ and $S_4$ ) and at fixed permeability $\kappa$, the estimate $\overline{D_e}$ increased slightly with the extracellular diffusion coefficient $D_e$, but the models compensated mainly by a smaller
$\overline{ICVF}$. 

Conversely, CEXI showed high sensitivity to $D_e$ changes with a limited impact on the estimates $\overline{R}$ and $\overline{ICVF}$. The CEXI $\overline{D_e}$ estimates matched the long-time limit of the $ADK_{ex}(t)$ in impermeable substrates derived as in Fig.~\ref{fig:signal_ADC_ADK_in_ex}, A,  supporting that the contribution of each compartment to the full signal could be disentangled by including exchange in the model. 

\subsubsection{Effect of the mean cell size on the negligible exchange assumption}

The substrate mean volume-weighted radius $R$ is an important property as it influences both the exchange time $\tau_{ex}$ \citep{Fieremans2010} and the characteristic diffusion time $t_c = \frac{R^2}{6D}$.
Recently, the effect of impermeable spherical cells were shown to contribute significantly to the diffusion signal at long diffusion time $(\Delta > 20ms)$ and pulse duration $(\delta > 1 ms)$ in tissue with an intrasoma volume fraction around 10\%-20\% \citep{Olesen2022}. For this range of volume fraction, \cite{Afzali2021} reported a lower bound of $3\mu m$ for the radius estimation using SANDI. Fig.~\ref{fig:model_single_sphere_allmethods}A suggested that SANDI can be used to estimate the volume-weighted radius $R$ down to $2\mu m$ in impermeable substrates with one population of spheres.  Interestingly, the MAE of SANDI on the cell size $\overline{R}$ was in the same range for all impermeable substrates. As a result, the relative error decreased from $10\%$ for $R_s=2\mu m$ down to $1\%$ for $R_s=5 \mu m$. The larger intracellular volume fraction $ICVF$ of the substrates might explain the better estimates of $\overline{R}$ in comparison to \cite{Afzali2021}. 

In permeable substrates, the relative error of the SANDI $\overline{R}$ also decreased with an increasing mean cell size. The smallest spheres were underestimated by more than $70\%$ (Fig.~\ref{fig:model_single_sphere_allmethods}A, $S_1$), possibly due to the fast exchange being already dominant in the signal. Because the relative error decrease with the mean cell size, the range of permeability $\kappa$ with an acceptable error broadened. For example, the relative error was under $20\%$ for all permeability in the bigger spheres ($S_4$), while this accuracy was reached at slower permeability in smaller cells ($\kappa < 10~\frac{\mu m}{s}$ in $S_2)$. The better estimation of the cell size $\overline{R}$ might be explained by the proportion of particles that did not encounter the membranes during simulation and, therefore, contributed to the signal as if they were diffusing in impermeable tissue. In larger cells, this proportion decreased and more particles met the SANDI assumptions, which agreed perfectly with the suggestion of using short diffusion time \citep{Palombo2020a}. The same effect was visible with VERDICT in the large cells ($R > 4~\mu m$ ). The minimal cell size $R$ for a small error was larger for VERDICT, but the error above this threshold was smaller with VERDICT than SANDI. This suggest that SANDI is probably more suitable in impermeable tissue with an unknown cell radius and a shorter exchange time, while VERDICT might output better estimation of large cells. In the presence of exchange, the CEXI estimate of the cell size $\overline{R}$ was again the most stable across permeability $\kappa$ and cell size $R$, with a maximal overestimation of $30\%$ when $R>2~\mu m$. 

Among all experiments, SANDI and VERDICT estimated the intracellular volume fraction $ICVF$ accurately in impermeable substrate with big cells only (Fig.~\ref{fig:model_single_sphere_allmethods}A,B, $S_3$, $S_4$). In permeable substrates, the estimate $\overline{ICVF}$ was nearly independent on the cell size, \ie~the estimates were close for different substrates with the same permeability and extracellular diffusion coefficient. Introducing exchange with CEXI improved significantly the estimates $\overline{ICVF}$ in the substrates with a cell size $R < 4 \mu m$, especially in the smallest sphere $S_1$ with a maximal relative error of $20\%$ for all $\kappa$ (Fig.~\ref{fig:model_single_sphere_allmethods}C). Conversely to VERDICT and SANDI, the relative error of CEXI on the intracellular volume fraction $ICVF$ increased with the cell size $R$, until reaching a value similar to SANDI and VERDICT in the worst case ($S_4$, $\kappa=50~\frac{\mu m}{s}$). 

For all methods and substrate properties, the quality of the results decreased in multiple-$R_s$ substrates (Fig.~\ref{fig:model_mul_sphere_allmethods}). Among the multiple-$R_s$ substrates, the estimates $\overline{R}$ of the volume-weighted mean cell size were notably better in the substrates with a smaller gap between the spheres ($M_1$ and $M_3$). Because the models were developed for the estimation of a single radius,  the $\overline{R}$ was expected to be close to the mean volume-weighted radius $R$. In spheres with very distinct sizes, the signals might be too different for this approximation to be valid, and a three-compartment model with exchange might by needed, at the cost of additional parameters. As soon as the membrane was permeable, SANDI and VERDICT estimates were not relevant anymore. Again, the intracellular volume fraction estimate $\overline{ICVF}$ of CEXI remained accurate in the substrates $M_1$ and $M_3$ at moderate permeability $\kappa < 10\frac{\mu m}{s}$, which was coherent with the validity of the exchange time estimations of Fig.~\ref{fig:cexi_tex}.  

\subsection{Permeability estimation with CEXI} \label{sec:discussion_perm}

Similarly to NEXI \citep{jelescu2022}, the permeability $\kappa$ was well estimated with multiple diffusion time $\Delta$ only (Fig.~\ref{fig:cexi_tex}). At low SNR, the MAE of $\overline{\kappa}$ remained acceptable contrary to the results presented in~\cite{jelescu2022}, possible due to the larger intracellular volume fraction $ICVF$. 

\begin{figure}[h!]
\begin{center}
\includegraphics[width=0.8\textwidth]{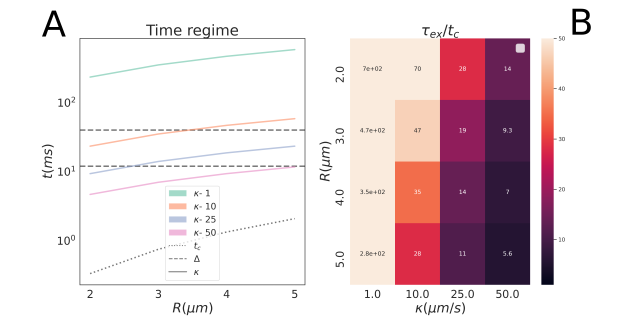}
\end{center}
\caption{\textbf{Time regime of the simulations.} (A) The full coloured lines show the evolution of the exchange time $\tau_{ex}$ with the mean sphere size $R_s$ and the permeability $\kappa$ (colour). The dotted black line shows the characteristic time $t_c = \frac{R_s^2}{6 D}$ of the substrates, and the dashed black lines show the shortest $\Delta_{min} = 12ms$ and longest  $\Delta_{max} = 40ms$ diffusion times. At short permeability $\kappa < 10 \frac{\mu m}{s}$, the exchange time was comparable to the diffusion time, and the effect of permeability competed with the diffusion inside compartments.  In the very permeable substrates $\kappa > 25 \frac{\mu m}{s}$, the compartments were already mixed at the shortest diffusion time, explaining the poorest results of CEXI in this range. (B)Ratio of the exchange time $\tau_{ex}$ and the characteristic time $t_c$. For all pairs $(\kappa - R_s)$, $\tau_{ex}$ exceeded $t_c$ by at least one order of magnitude.}\label{fig:cexi_time_regime}
\end{figure}

CEXI assumes that diffusion is barrier-limited, \ie~ diffusion inside each compartment is Gaussian, and the exchange time $\tau_{ex} = \frac{R}{3\kappa} (1 - ICVF ) $ \citep{Fieremans2010} is much longer than the characteristic diffusion time $t_c = \frac{R^2}{6D_i}$ of the medium. The first condition was always satisfied because the longest characteristic time ($t_c = 2 ms$, $S_4$) was always shorter than the shortest diffusion time ($\Delta = 12 ms$) (Fig.~\ref{fig:cexi_time_regime}B). Conversely, the second condition of the barrier-limited regime was more complicated to satisfy as the permeability $\kappa$ increased (Fig.~\ref{fig:cexi_time_regime}A). At slow permeability ($\kappa < 10 \frac{\mu m}{s}$), the exchange time $\tau_{ex}$ was longer than the characteristic time $t_c$ and the diffusion time $\Delta$ for all substrates. At intermediate permeability ($\kappa \in \left[10, 25 \right] \frac{\mu m }{s}$), $\tau_{ex}$ and $\Delta$ had similar order of magnitude for all substrates, which explains the stability of the NEXI estimates in this range of permeability $\kappa$. Finally, in the substrates with a fast permeability ($\kappa > 25 \frac{\mu m}{s}$), CEXI became sensitive to the exchange in the biggest spheres only because the exchange time $\tau_{ex}$ was too short in the small cells. Consequently, the error of $\overline{\kappa}$  decreased with an increasing cell size $R$. 

\subsection{Recommendations}

In the light of the results discussed in Section~\ref{sec:discussion_model}, SANDI and VERDICT remains the best options in impermeable tissue, with a preference for SANDI when the substrate properties are not known. However, the volume-weighted mean cell size $\overline{R}$ and the intracellular volume fraction $\overline{ICVF}$ estimates deteriorated quickly with the permeability. SANDI and VERDICT compensated for the effects of the permeability by an underestimation of $ICVF$ and an overestimation of $R$. In permeable tissue, CEXI provided more robust estimates in the investigated range of parameters. In the biggest sphere or at high permeability, estimation remained challenging even with CEXI because the exchange was too slow or too fast respectively for the range of diffusion time considered in this work. Future work should determine if the microstructure properties in this area could be more accurately estimated by optimizing the diffusion times and the diffusion gradient amplitude of the PGSE sequence.

Results described in Section~\ref{sec:discussion_perm} shows that accurate permeability estimation is possible with the model including exchange and an appropriate acquisition protocol. For that, the order of magnitude of both the characteristic size of the microstructure and the permeability must be known. Short diffusion times must be used in small biological structures or at high permeability, \eg~tumors, and longer diffusion times in larger biological structures, \eg~somas, and impermeable membranes, \eg~axons.  Future work using real data should confirm that PGSE with specific diffusion times could be used to probe membrane permeability and cell size at different time scales.

\section{Conclusion}\label{sec::conclusion}
In non-white matter tissue, the assumption of negligible exchange has been justified by short diffusion times. This work showed, with simulations in numerical substrates of tumors, that the exchange between the intracellular and the extracellular space cannot be neglected in permeable tissue with a conventional PGSE sequence. Additionally, the inherent bias in the estimates of the compartmentalized models for impermeable tissue was amplified in permeable tissue, even with very low permeability. As an alternative, a two-compartment model of tumors with the water exchange between spherical cells and the extracellular space was described to estimate the exchange time and the cell size simultaneously. Limitations of this model in substrates with multiple cell populations or in the presence of fast exchange were highlighted, justifying the need for a more sophisticated model of realistic tissue.


\section*{Conflict of Interest Statement}
The authors declare that the research was conducted in the absence of any commercial or financial relationships that could be construed as a potential conflict of interest.

\section*{Author Contributions}
RG: Methodology, Coding, Simulations, Analysis, Writing, Visualization,
JLVH: Discussion about substrate generation and the choice of the simulations parameters, Writing - Review,
EJC-R: Methodology, Writing-Review and Editing,
IOJ: Discussion about compartmentalized model and mesoscopic disorded theory, Writing-Review and Editing,
GG: Methodology, Experimental design, Writing-Review and Editing,
JR-P: Methodology, Experimental design, Discussion about simulations, Active contribution to the analysis of the results, Supervision, Writing - Review and Editing,
J-PT: Supervision, Funding, Writing-Review 

\section*{Funding}
This work is supported by the Swiss National Science Foundation under grants 205320\_175974 and 205320\_204097.

\section*{Acknowledgments}
We acknowledge access to the facilities and expertise of the CIBM Center for Biomedical Imaging, a Swiss research center of excellence founded and supported by Lausanne University Hospital (CHUV), University of Lausanne (UNIL), Ecole polytechnique fédérale de Lausanne (EPFL), University of Geneva (UNIGE) and Geneva University Hospitals (HUG). Erick J. Canales-Rodríguez was supported by the Swiss National Science Foundation (Ambizione grant PZ00P2\_185814). We thank Jenifer Miehlbradt for her advice on the writing of this manuscript.

\clearpage

\clearpage
\begin{table}[h!]
    \centering
    \begin{tabular}{|c|c|c|c|c|c|c|c|}
    \hline
         Substrate      &  $S_1$ &  $S_2$ &  $S_3$ &  $S_4$ &  $M_1$ &  $M_2$ &  $M_3$\\\hline
         Voxel side ($\mu m$)  & \multicolumn{7}{c|}{$100$} \\\hline
         ICVF           & \multicolumn{7}{c|}{$0.65$}\\\hline
         $R_s$ ($\mu m$)  & 2     & 3     & 4     & 5     & 1(60\%)-3 (40\%) & 1(60\%)-5(40\%) & 3(60\%)-5 (40\%)\\\hline
         $R$ ($\mu m$)    & 2.1   & 3.1   & 4.0   & 5.1   & 1.8   & 2.9   & 3.7\\\hline
    \end{tabular}
    \caption{Properties of the substrates. The voxel side length and the intracellular volume fraction (ICVF) were the same for all substrates. $R_s$ was the mean cell radius of the sphere populations, and $R$ was the volume-weighted mean cell radius. Signals of the first experiment ($E_1$) were generated with the substrate $M_1$. Signals of the second experiment ($E_2$) were generated with all substrates.}\label{tab:substrate}
\end{table}

\begin{table}[h!]
    \centering
    \begin{tabular}{|c|c|c|c|c|c|}
    \hline
         Experiments      &  $D_{e,0} (\frac{\mu m^2}{s})$ &  $D_{i,0} (\frac{\mu m^2}{s})$ &  $\kappa (\frac{\mu m}{s})$ &  $\rho (\frac{part}{\mu m^3})$ & $\delta t (\mu s)$\\\hline
         $E_1$  & 0.5, 1 ,2     & 1, 2  & 0, 5, 10, 50, 100     & 2   & 5\\\hline
         $E_2$  & 0.5, 1, 2     & 2     & 0, 10, 25, 50         & 0.5 & 5     \\\hline
    \end{tabular}
    \caption{Simulation parameters of the experiments. The first experiment ($E_1$) was generated with more particles and a wider permeability range than the second experiment ($E_2$).}\label{tab:experiment}
\end{table}

\begin{table}[h!]
    \centering
    \begin{tabular}{|c|c|c|c|c|c|c|}
    \hline
         Sequence      &  $\Delta (ms)$ &  $\delta (ms)$ &  $TE (ms)$ &  $b (\frac{ms}{\mu m^2})$ & $g (\frac{mT}{m})$     & N\\\hline
         $E_1$      & 12, 20, 30, 40    & 4.5       & 50    & 1, 2.5, 4, 5.5, 7                      & -         & 24 \\\hline
         SANDI      & 11, 20            & 3         & 30    &  0, 1, 2.5, 3, 4, 5.5, 7, 8.5, 10     & -         & 24 \\\hline
         VERDICT    & 10, 20, 30, 40    & 3, 10     & 50    & -                                     & 40-400    & 3 \\\hline
         NEXI       & 12, 20, 30, 40    & 4.5       & 50    & 1,   2.5, 4, 5.5, 7                      & -         & 24 \\\hline
    \end{tabular}
    \caption{Parameters of the PGSE sequences of the experiment described in Section \ref{sec:method_model}. Signals generated with the SANDI, VERDICT and NEXI protocols were fitted to their respective models. NEXI protocol was used for performance comparison across models.}\label{tab:pgse}
\end{table}

\clearpage
\bibliographystyle{preprint} 
\bibliography{biblio}

\end{document}